\documentstyle[12pt,epsf,epsfig,wrapfig]{article}

\begin{document}

\begin{center}

{\bfseries EXOTIC STATE SEARCHES AT THE SPRING-8:
\\
OBSERVATION OF A PENTAQUARK STATE $\Theta^+$ BARYON}

\vskip 5mm

Yuji Ohashi on behalf of the LEPS collaboration
\vskip 5mm

{\small

 {\it
 SPring-8\\
Japan Synchrotron Radiation Research Institute\\
Mikazuki, Sayo, Hyogo, 679-5198, Japan\\
}
\vskip 5mm
 {\it

E-mail: ohashi@spring8.or.jp

}}

\end{center}

\vskip 5mm

\begin{abstract}

A narrow resonance peak was observed at $1.54 \pm 0.01$ GeV/$c^2$ as an exotic baryon with strangeness $S = +1$ in the $\gamma + n \rightarrow K^+ + K^- + n$ reaction on $^{12}$C. The new state have a width smaller than $0.025$ GeV/$c^2$ and a Gaussian significance of $4.6\sigma$. It can 
not be interpreted as three quark state and is consistent with the lightest member of an antidecuplet of baryons predicted by the chiral soliton model.

\end{abstract}

\vskip 8mm
\vspace{4mm}
\begin{center}
{\large \bf INTRODUCTION}
\end{center}

The constituent quark model has been remarkably successful to 
describe most of the experimentally observed mesons and baryons. 
Beyond conventional hadrons, exotic states such as glueballs, hybrid 
mesons, and other multi-quark states are not prohibited in QCD.
There is a different approach by the Skyrme model~\cite{s1} that is 
originated in the same QCD.
In this direction, following the pioneering works~\cite{s2,s3}, 
Diakonov, Petrov and Polyakov (DPP) proposed a possible 
existence of the $S = +1\  J^P = 1/2^+$ resonance at $1.53$ 
GeV/$c^2$ with a width less than $0.015$ GeV/$c^2$ using the 
chiral soliton model~\cite{dpp}. The $\Theta^+$ (originally denoted 
the $Z^+$) is the lightest member of the anti-decuplet which appears 
as the third rotational excitation states. Their model simply assumes the 
relatively well established nucleon resonance $N (1710, 1/2^+) $ as 
a nucleon-like member of the anti-decuplet. The mass formula for the 
members of the antidecuplet is proposed as
\begin{eqnarray}
M = [1.89 -Y \times 0.18\ ]  \ GeV/c^2
\end{eqnarray}
where Y is a hypercharge.
 
Motivated by the resonance-like structure appeared 
around $K^+$ momentum of $1$ GeV/$c$ in 
both $K^+p$ and $K^+n$ total cross-sections, experimental 
studies on the $K^+N$ reactions started from 1960's. It is to be noted 
that most of the searches were performed in the mass range 
between 1.74 and 2.16 GeV/$c^2$. Although the possible $Z^*$ baryon 
resonances were reviewed in the 1986 baryon listings 
by the PDG~\cite{pdg}, evidence of their existence was concluded as poor.

One of our experimental goal is to search various kinds of 
exotic states through photo-nuclear reactions. The present 
experiment was motivated in part by the DPP prediction mentioned 
above, and the experimental result was published in the 
article~\cite{nprl}. We describe this paper including some issues 
that were not mentioned at the workshop or appeared afterward.

\vspace{6mm}
\begin{center}
{\large \bf EXPERIMENTAL METHOD}
\end{center}

\begin{wrapfigure}{R}{7.5cm}

\mbox{\epsfig{figure=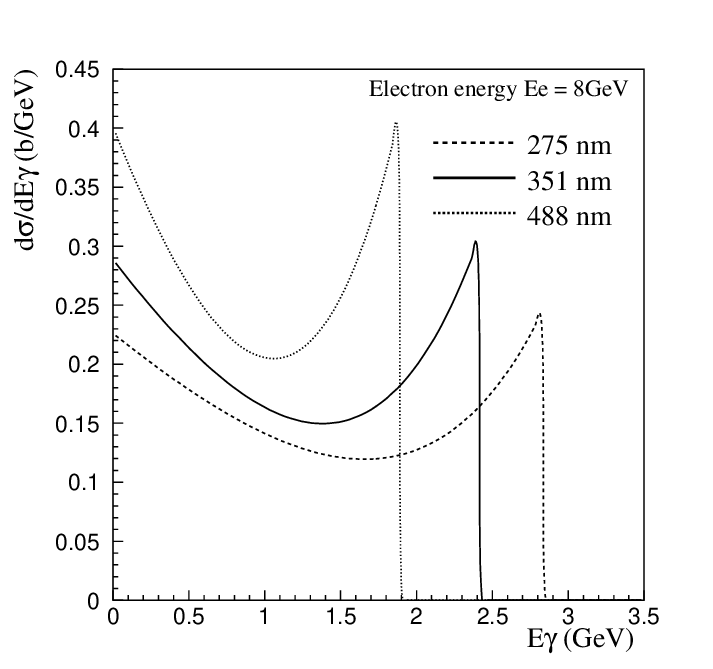,width=7cm}}

{\small{Figure 1.} Energy spectrum of a Laser-Electron Photon beam.\\
\ }

\end{wrapfigure}

The experiment was carried out at the Laser-Electron Photon facility 
at the SPring-8 (LEPS)~\cite{nnpa}. 
Photon beam is produced by Compton scattering of laser photons 
on the circulating 8 GeV electrons and beam energy is tagged by 
analyzing scattered electron momentum. 
The maximum photon energy is determined by the wave length 
of the laser, and it is 2.4 GeV when a 351 $n$m Ar laser is used. 

Energy spectrum of a photon beam is relatively flat as shown 
in Fig. 1 and photons are highly polarized when laser photons 
are polarized. Un-collimated beam size in standard 
deviation($\sigma$) is 3 mm vertcally and 8 mm horizontally 
at the target position about 70 m downstream of the Compton scattering region.
Typical tagged photon flux above 1.5 GeV was $\sim1 \times 10^6$ per second.

\begin{wrapfigure}{R}{9.5cm}
\vspace{-3mm}
\hspace{2mm}
\mbox{\epsfig{figure=yof2.epsf,width=8.1cm}}

{\small{Figure 2.} The LEPS detector setup.
\vspace{2mm}
}

\end{wrapfigure}

Experimental setup was designed for the study of $\phi$ meson 
production from protons. 
Photons hit a liquid hydrogen target (LH$_2$) of 5 cm thick after 
passing through a charged particle veto counter (VC).
The LEPS detector consisted of a start trigger counter (SC), a 
silica aerogel ${\rm\check{C}}$erenkov counter (AC), a silicon-strip 
vertex detector (SSD), an upstream drift chamber, a 0.7T dipole magnet, 
two sets of downstream drift chambers, and a time-of-flight scintillator 
array(TOF) as shown in Fig. 2. 
The detector acceptance covered forward angle of $\pm$0.4 
and $\pm$0.2 rad in the horizontal and vertical directions, respectively. 
Typical detector resolutions in $\sigma$ for tagged $E_{\gamma}$, 
momentum, time-of-flight, and mass (momentum dependent) were 
0.015 GeV/$c^2$, 0.006 GeV/c at 1 GeV/c, 150 $p$sec for a flight length 
of  $\sim 4$ m from the target to the TOF,  and 0.01 GeV/$c^2$ 
for 1 GeV/c kaon, respectively. Particle identification was applied 
within 3 $\sigma$ in this analysis, and its effectiveness can be 
viewed in the mass spectrum in Fig. 3.

\setcounter{figure}{2}

\begin{figure}[t]
\vspace{-5mm}
\centerline{\epsfysize=2.6in\epsfbox{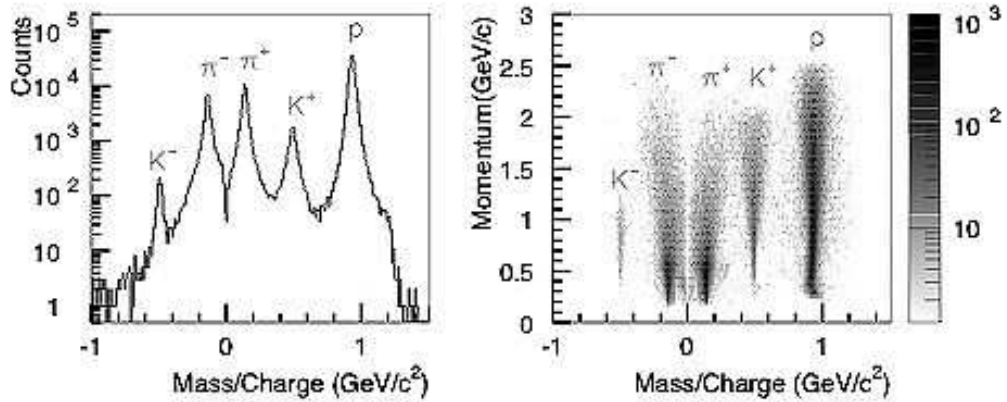}}
\vspace{-10mm}
\caption{The Mass distribution.
\label{fig:mass2} }
\end{figure}

$e^+e^-$ pairs produced in the very forward angles were blocked by 
horizontal lead bars which were set in the median plane inside the magnet gap. 
The trigger simply required a hadron event by applying a tagging 
counter hit, no charged particle before the target, charged particles 
after the target, no signal in the AC, and at least one hit on the TOF wall.
AC fires for electrons and positrons, and for pions with momentum 
higher than 0.6 GeV/c by the choice of a refractive index of 1.03.
A typical trigger rate was about 20 events per second.
The data taking was carried out during December, 2000 to June, 2001. 
\vspace{5mm}
\begin{center}
{\large \bf DATA ANALYSIS}
\end{center}

\begin{wrapfigure}{R}{6.5cm}
\vspace{-6mm}
\hspace{3mm}
\mbox{\epsfig{figure=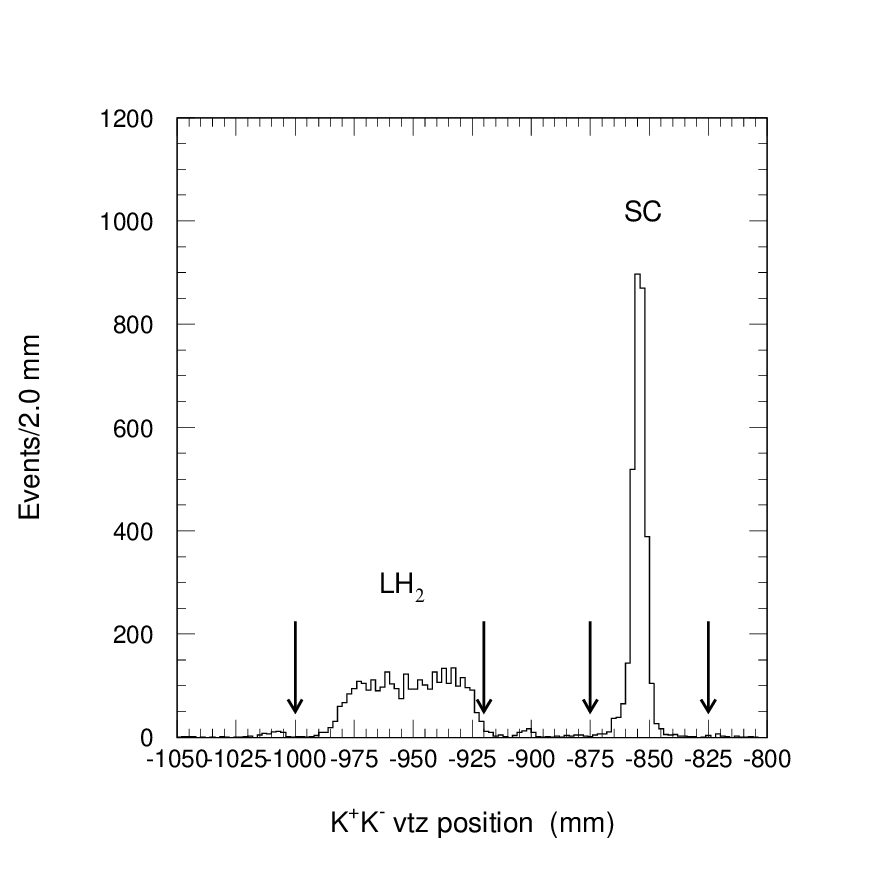,width=5.5cm}}

{\small{Figure 4.} Vertex position for $K^+K^-$ events along the 
photon beam direction. Arrows indicate the cut position for SC 
or LH$_2$ events. SC events were used for exotic state search.}

\end{wrapfigure}

From the total amount of $4.3 \times 10^7$ triggers, $8.0 \times 10^3$ 
events with a $K^+K^-$ pair were reconstructed. Data reduction was carried out 
in the following scheme.\\
\ 
\\
(1) Vertex position cut

The SC, which was used as a neutron target for the search 
of  $\Theta^+$, was a 0.5 cm thick plastic scintillator (polystyrene) 
and was placed at 9.5 cm downstream of LH$_2$. 
$K^+K^-$ pair events from the SC were identified by selecting 
vertex position along beam direction. Events that originate from 
a reaction in the SC were clearly seen in Fig. 4.
Events from maximum photon energy region above 2.35 GeV 
were omitted in consideration of increasing background 
contribution to the signal with increasing energy.
These cuts reduced the data sample to $3.2 \times 10^3 K^+K^-$ events.
\\
(2) Missing mass cut
\begin{wrapfigure}{R}{7cm}
\vspace{-5mm}
\hspace{3mm}
\mbox{\epsfig{figure=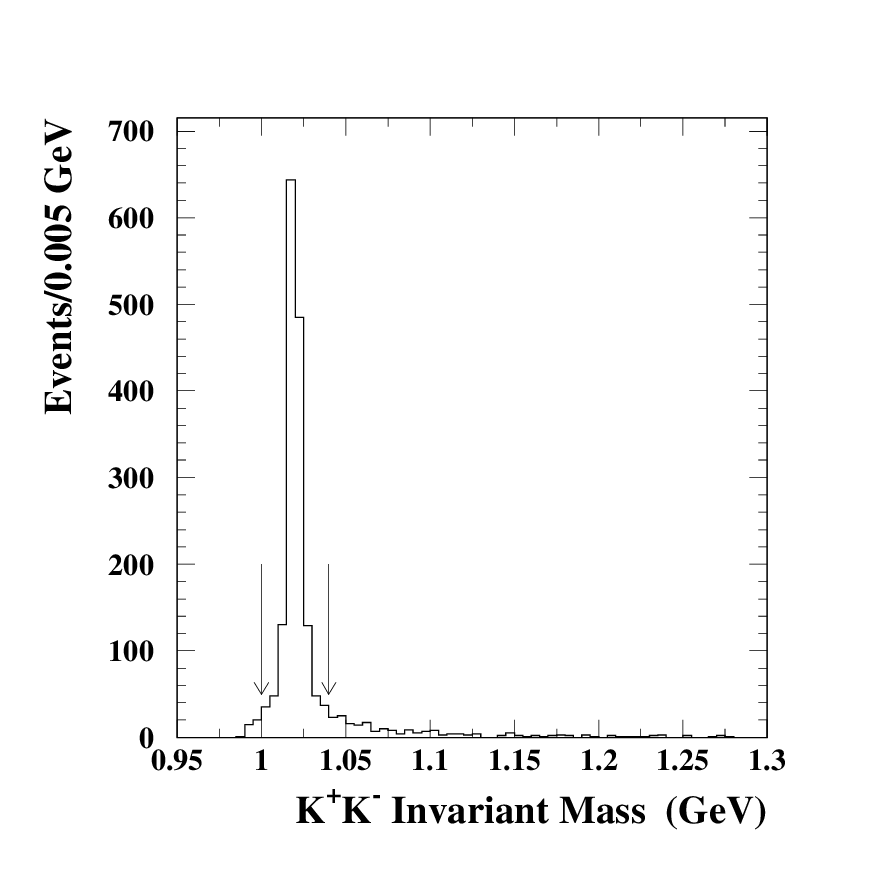,width=6.5cm}}

{\small{Figure 5.} Invariant $K^+K^-$ mass distributions for the SC events. 
$\phi$ meson contributions were cut for exotic state search as indicated by arrows.}

\end{wrapfigure}

In the reaction $\gamma N \rightarrow K^+ K^- X$, $M_X$ was 
required to be nucleon and the rest of carbon nucleus be spectator.
The missing mass $MM_{\gamma K^+K^-}$ in the above reaction 
was calculated by assuming a qusi-free $\gamma N$ scattering. 
A total of $1.8 \times 10^3$ events survived after applying 
$0.90 < MM_{\gamma K^+K^-} < 0.98$ GeV/$c^2$.
\\
(3) $\phi$ meson cut

Most of the remaining $K^+K^-$ events (~85\%), which were 
from $\phi$ meson photoproduction and were uninteresting 
events in this report, were cut by applying 
$1.00 < M(K^+K^-) < 1.04$ GeV/$c^{2}$ as in Fig. 5.
\\
(4) Proton cut

$K^+K^-$ events, of which reactions were originated from 
protons, were eliminated by requiring SSD hit. The direction 
and momentum of the nucleon in the final state was calculated 
from the $K^+$ and $K^-$ momenta. Events were rejected if 
the recoiled nucleon was out of the SSD acceptance or was 
the momentum smaller than 0.35 GeV/c since an uncertainty 
in the calculation becomes large for low momentum case.
108 events were eliminated by requiring that the hit position 
in the SSD agreed with the expected hit position within 45 mm. 
This distance was chosen to take the Fermi motion in $^{12}$C into account.  
A total of 109 events satisfied all the selection criteria ("signal sample").
\begin{wrapfigure}{R}{7cm}
\vspace{-9mm}
\hspace{3mm}
\mbox{\epsfig{figure=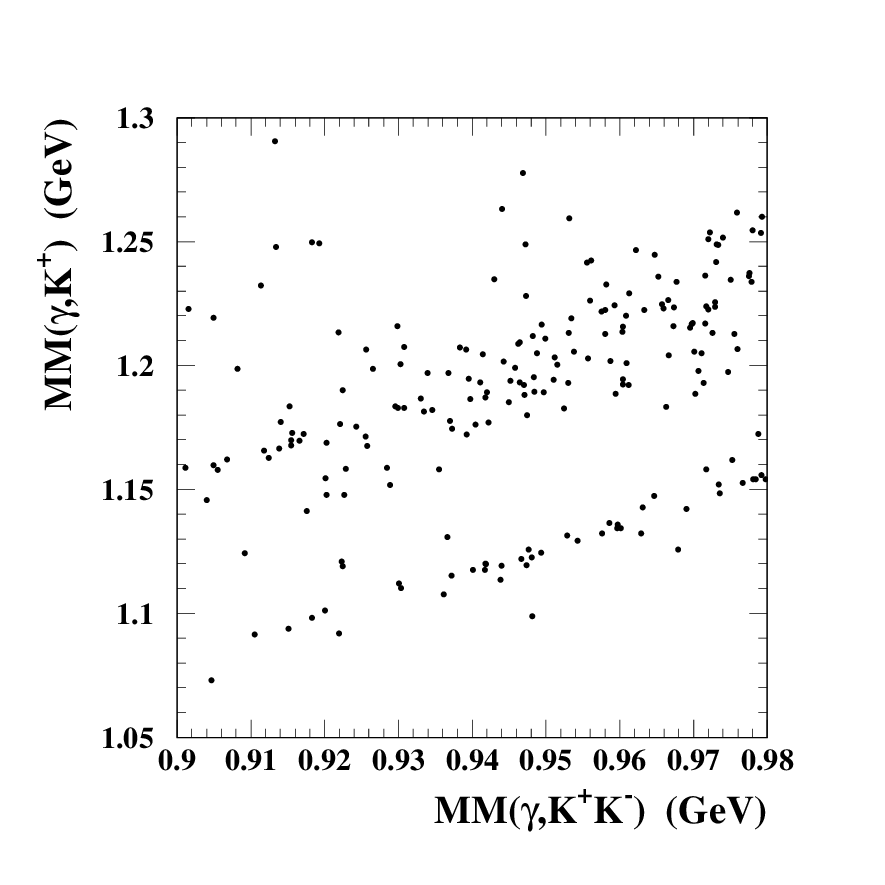,width=6.5cm}}

{\small{Figure 6.} The scatter plot of $MM_{\gamma K^+}$ 
$vs.$ $MM_{\gamma K^+\pi^-}$ for $K^+\pi^-$ photoproductions in the SC.
\vspace{2mm}
}

\end{wrapfigure}

A raw missing mass calculation done above was obtained 
without considering the Fermi motion in $^{12}$C. To evaluate 
this effect, well known process 
$\gamma n \rightarrow K^+\Sigma^- \rightarrow K^+\pi^-n$ was 
studied. Clear correlations are seen in the missing mass plots Fig. 6, where 
$MM_{\gamma K^+}$ and $MM_{\gamma K^+\pi^-}$ were obtained for 
the $\gamma N \rightarrow K^+X$ and $\gamma N \rightarrow K^+\pi^-N$ 
channels by assuming the nucleons at rest. This is understood by the 
fact that the nucleons in the two channels are identical. Considering the 
above missing mass correlation and setting the nucleon mass to $M_N$, 
the Fermi motion corrected missing mass $MM_{\gamma K^+}^c$ is obtained as
\begin{eqnarray}
MM_{\gamma K^+}^c = MM_{\gamma K^+} - MM_{\gamma K^+\pi^-} + M_N.
\end{eqnarray}
Eq. (2) compensates not only the Fermi motion but also the experimental 
resolutions and the binding energy of the nucleon in $^{12}$C. The Fermi 
motion effect is the major contribution in the analysis. 
Although there is only a smeared distribution in the raw 
$MM_{\gamma K^+}$ (dashed line in Fig.7), the 
$\Lambda$ and the $\Sigma^-$ peaks are separated after 
the correction (solid line in Fig.7). 
Note that this correction is not a good approximation for events 
with $MM_{\gamma K^+\pi^-}$ far from the nucleon rest mass, 
and is good in case of a decay with a small Q value. 
\setcounter{figure}{6}
\begin{figure}[t]
\centerline{\epsfxsize=2.7in\epsfbox{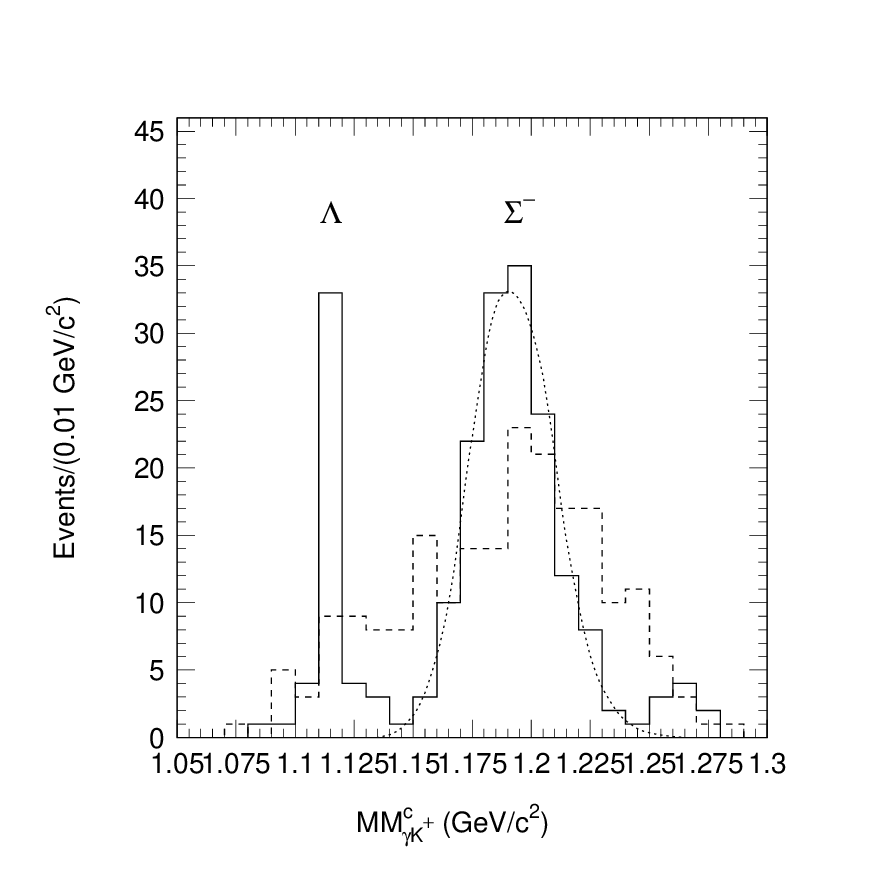}}
\vspace{-5mm}
\caption{The corrected missing mass distribution 
$MM_{\gamma K^+}^c$ [Eq.(2)] for the $K^+\pi^-$ events from 
the SC (solid line) and for Monte Carlo events for the 
$\gamma n \rightarrow K^+\Sigma^-$ channel (dotted curve). 
The dashed line shows the missing mass distribution 
$MM_{\gamma K^+}$ which was obtained from the projection of the 
events in Fig. 6 on to the vertical axis without the Fermi-motion correction. 
\label{fig:sigma} }
\end{figure}
The latter effect is seen in the small width for the 
$\Lambda$ and the large width for the $\Sigma^-$ because of the 
imperfection of the correction due to a large Q value for the $\Sigma^-$.

The corrected missing mass $MM_{\gamma K^\pm}^c$  
for $K^+K^-$ events is given as
\begin{eqnarray}
MM_{\gamma K^\pm}^c = MM_{\gamma K^\pm} - MM_{\gamma K^+K^-} + M_N.
\end{eqnarray}

Fig. 8 shows no obvious peaks in a $MM_{\gamma K^+}^c$ 
distribution of the signal sample (109 events) that satisfied all 
the selection criteria (dotted line), and shows a clear peak due to the 
$\gamma p \rightarrow K^+\Lambda$(1520) $\rightarrow K^+K^-p$ 
process in the 108 events that agreed the proton hit in the SSD (solid line).
Effectiveness of a proton cut is verified that the $\Lambda$(1520) 
peak does not show up in the dotted line plot. 

\vspace{-4mm}
\begin{figure}[h]
\centerline{\epsfxsize=2.6in\epsfbox{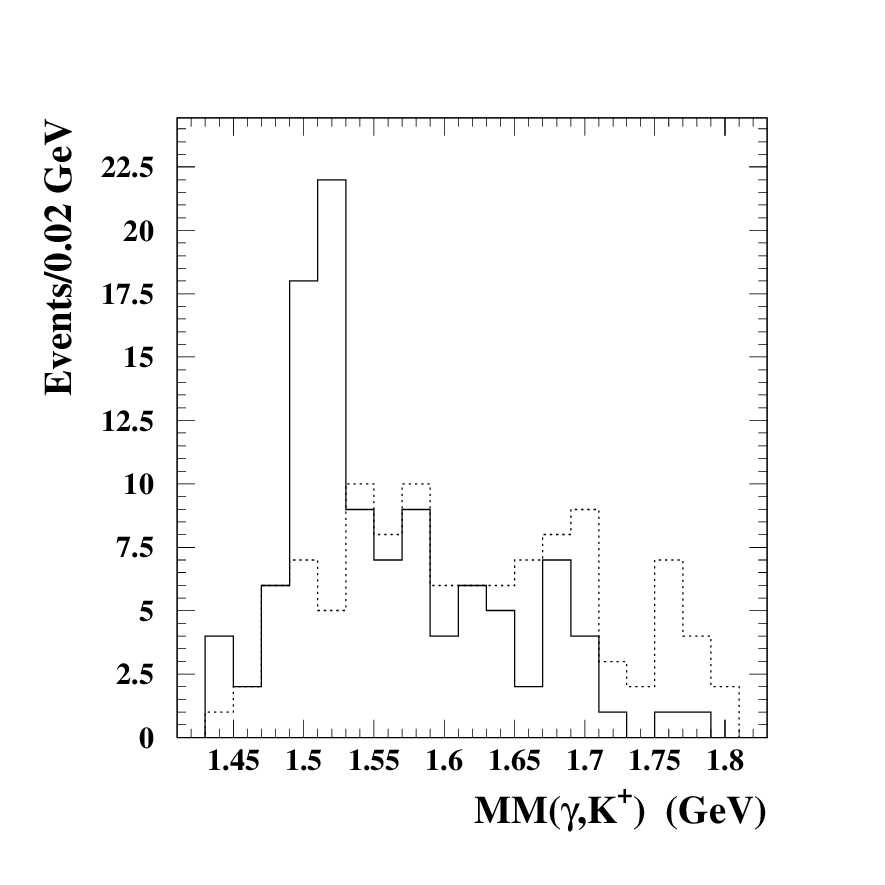}}
\vspace{-5mm}
\caption{The corrected missing mass distribution 
$MM_{\gamma K^+}^c$ [Eq.(3)] for the $K^+K^-$ productions for 
the signal sample (dotted line) and for events from the SC with a 
proton hit in the SSD (solid line).
\label{fig:sigma} }
\end{figure}

Fig. 9 shows the $MM_{\gamma K^-}^c$ distribution of the signal sample. 
A prominent peak at 1.54 GeV/$c^2$ is observed with 36 events in the 
peak region between 1.51 and 1.57 GeV/$c^2$. 
To estimate the background due to the non-resonant $K^+K^-$ 
production in the peak region defined above, 
the missing mass distribution of the signal sample in the 
region above 1.59 GeV/$c^2$ was fitted by a distribution of events 
from the LH$_2$. LH$_2$ sample required the same selection criteria 
except for the shifted $vtz$ window and a proton hit in the SSD.
Since the $\Lambda$(1520) contribution was removed from the signal 
sample, it was removed from LH$_2$ sample for 
$1.51 \leq MM_{\gamma K^+}^c < 1.53$ GeV/$c^2$.
The best fit with a $\chi^2$ of 7.2 for 8 degrees of freedom is obtained 
with a scale factor of 0.2 as shown with a dotted line in Fig. 9.
Resulting background level in the peak region is estimated to be 17.0 
$\pm$ 2.2 $\pm$ 1.8, 
where the first uncertainty is the error in the fitting in the region above 
1.59 GeV/$c^2$ and the second is a statistical uncertainty in the peak region. 
The combined uncertainty of the background level is $\pm$2.8. 
The estimated number of the signal events after subtracting 
background is 19.0 $\pm$ 2.8, which corresponds to a 
Gaussian significance of $4.6_{-1.0}^{+1.2} \sigma$ 
(19.0/$\sqrt{17.0} $). 
The signal region between 1.47 and 1.61 GeV/$c^2$ after background 
subtraction was compared with Monte Carlo simulations assuming a 
Breit-Wigner function for a resonance shape.
The best fit was 1.54 $\pm$ 0.01 (statistical) GeV/$c^2$. 
The systematic error was speculated to 0.005 GeV/$c^2$ by referring the 
known particle $\Sigma^-$, of which peak position was 0.005 GeV/$c^2$ 
smaller than the PDG value of 1.197 GeV/$c^2$. The width $\Gamma$ 
of the resonance cannot be determined by the fitting since the zero 
width gives the minimum $\chi^2$ of 1.6 for 4 degrees of freedom. 
The upper limit for the width was determined to be 0.025 GeV/$c^2$ 
with a confidence level of 90\%.
\vspace{-8mm}
\begin{figure}[h]
\centerline{\epsfxsize=4.2in\epsfbox{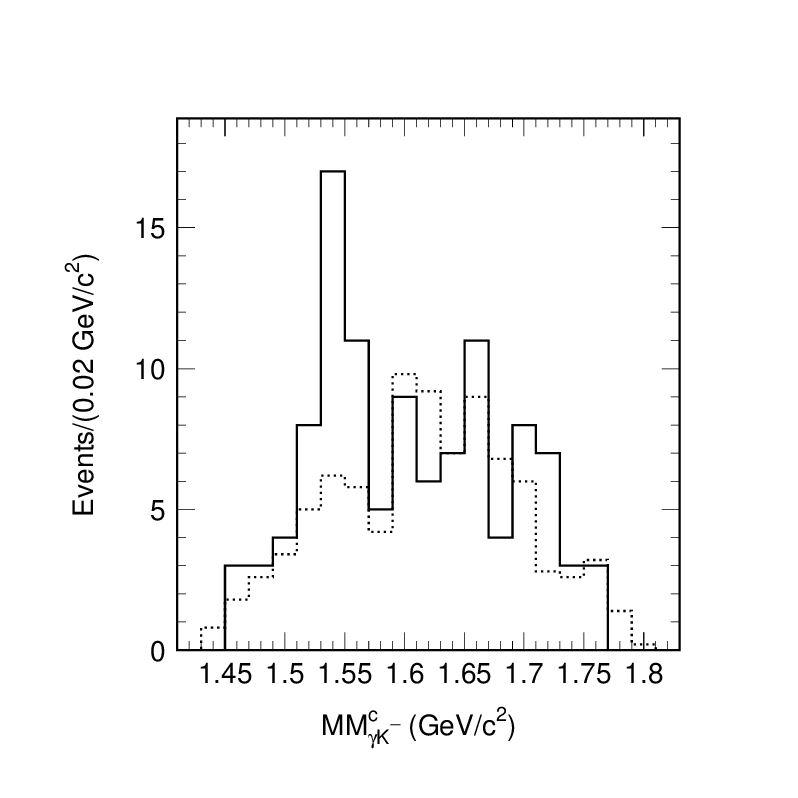}}
\vspace{0mm}
\caption{The corrected missing mass distribution $MM_{\gamma K^-}^c$ 
[Eq.(3)] for the signal sample (solid line) and for events from the LH$_2$ 
(dotted line)  normalized by a hit in the region above 1.59 $GeV/c^2$.
\label{fig:sigma} }
\end{figure}
\\

To make sure that the observed peak is not due to a fake one, the mass 
cut on one of the kaons was tightened from 3 $\sigma$ to 2 $\sigma$, 
and also the kaon mass was intentionally assigned for the pions in the 
missing mass calculations. The peak was not affected in both cases, 
and it was also confirmed that the peak is not due to the effect of the 
tails of the $\phi$ mesons.

\vspace{6mm}
\begin{center}
{\large \bf DISCUSSION}
\end{center}

After we announced the evidence of $\Theta^+$, significant deal of 
activities have been reported in both experimental and theoretical fields.
Because of the peculiar nature and relatively small statistical 
significance of the new state, experimental confirmation was seriously 
awaited. Several experimental results appeared in succession from the 
DIANA collaboration at the ITEP~\cite{diana}, CLAS collaboration at the 
JLab~\cite{clas}, SAPHIR collaboration at the ELSA~\cite{saphir}, and a 
neutrino experiment~\cite{nu}. in all cases the mass is near 1.54 GeV/$c^2$ 
and the width is smaller than 0.025 GeV/$c^2$. There are informal announces 
of again consistent observations of $\Theta^+$ from HERMES and ZEUS. 
It is instructive that the DIANA collaboration reanalyzed almost 15 years 
old data which was obtained in the direct $s$-channel $K^+N$ 
measurements with $X_e$ bubble chamber. 

The most surprising point was the narrowness of the $\Theta^+$ 
width which is claimed to be smaller than 0.009 GeV/$c^2$~\cite{diana}. 
Recent partial wave analysis on $K^+N$ elastic scattering by Arndt, 
Strakovsky and Workman indicated that the width of $\Theta^+$ is likely 
smaller than 0.001 GeV/$c^2$~\cite{psa}. One of the speculation required 
$\Theta^+$ to be isotensor, and the isospin violating decay causes 
the narrow width~\cite{isot}. This explanation seems unlike if isospin 
of $\Theta^+$ is confirmed to be zero~\cite{saphir,iso0}. 

In fact we only know the $\Theta^+$ mass but not other quantum 
numbers. Spin and parity of $\Theta^+$ can be determined in 
principle by measuring the decay angular distribution and nucleon 
polarization as in the textbook. This is extremely difficult to do 
experimentally in straightforward way. To determine quantum 
numbers is our crucial task now on~\cite{qn1,qn2,qn3}. 

It is natural to ask whether $\Theta^+$ is a resonance or a 
baryon-meson molecular like state. It seems unusual to 
assume $\Theta^+$ to be a baryon-meson molecular state 
because of its narrow width according to Jaffe and Wilczek (JW)~\cite{jw}.

A discovery of the $\Theta^+$ was certainly guided by the prediction 
based on the chiral soliton model by DPP. On the other hand quark 
model approaches have been reported recently with highly correlated 
diquak-diquark~\cite{jw} and diquark-triquark~\cite{kl} schemes. If $\Theta^+$ 
mass is set to experimentally measured value of 1.54 GeV/$c^2$ in 
each model, expected exotic $\Xi$ mass is quite different between 
correlated quark model and chiral soliton model predictions. 
Exotic $\Xi$ mass is at 1.75 GeV/$c^2$ by a quark model of 
JW, although it is at 2.07 GeV/$c^2$ from eq.(1) with S = -2 by 
a chiral soliton model by DDP. DDP predicts the width to be 
greater than 0.14 GeV/$c^2$ in addition.
Recently reported observation of exotic $\Xi$ by NA49 collaboration 
indicates that the mass is at 1.862 GeV/$c^2$ with a width narrower 
than 0.018 GeV/$c^2$~\cite{exi}. The measured mass does not fit in 
any predictions, and Diakonov and Petrov proposes a revised model~\cite{dp}. 
There have been significant discoveries recently in meson spectroscopy 
field from the B-collider experiments. Unusually narrow meson states were 
observed in the BABAR, BELLE and other detectors~\cite{ds}. 
These states may be attributed to exotic tetraquark states. Further 
researches on the $\Theta^+$ and exotic $\Xi$ baryons together 
with the exotic mesons will boost in understanding the nature of the exotic states.

\vspace{6mm}
\begin{center}
{\large \bf CONCLUSION}
\end{center}

The LEPS measurement provides the first evidence for the existence of a 
S = +1 narrow baryon resonance at 1.54 $\pm$ 0.01 GeV/$c^2$ in the 
$K^-$ missing mass spectrum of the $\gamma n \rightarrow K^+K^-n$ 
reaction on $^{12}$C. The Gaussian significance is estimated to be 
4.6 $\sigma$ and the width is smaller than 0.025 GeV/$c^2$. The result 
is surprisingly coincided with the DPP's prediction.
This state is certainly an exotic baryon with a quark configuration of $uudd\bar{s}$.

\vspace{6mm}
\begin{center}
{\large \bf ACKNOWLEDGMENTS}
\end{center}

I am grateful to the organizers of Spin2003 for their invitation. I would like 
to thank Dr. Atsushi Hosaka for stimulating discussions on 
exotic states. This research was supported in part by the Ministry of 
Education, Science, Sports and Culture of Japan, by the National Science 
Council of Republic of China (Taiwan), and by KOSEF of Republic of Korea.


\vspace{6mm}
\begin{thebibliography}{99}
\bibitem{s1}

T. H. R. Skyrme, Nucl. Phys. {\bfseries 31}, 556 (1962).

\bibitem{s2}

M. Chemtob, Nucl. Phys. {\bfseries B256}, 600 (1985).

\bibitem{s3}

H. Walliser, Nucl. Phys. {\bfseries A548}, 649 (1992).

\bibitem{dpp}

D. Diakonov, V. Petrov, and M. Polyakov, Z. Phys. {\bfseries A359}, 305  (1997).

\bibitem{pdg}

Particle Data Group, Phys. Lett. {\bfseries 170B}, 289 (1986).

\bibitem{nprl}

T. Nakano et al., LEPS collaboration, Phys. Rev. Lett. {\bfseries 91}, 012002 (2003).

\bibitem{nnpa}

T. Nakano et al., LEPS collaboration, Nucl. Phys. {\bfseries A684}, 71c (2001).

\bibitem{diana}

V. V. Barmin et al., DIANA collaboration, Yad. Fiz., {\bfseries 66}, issue 9 (2003); hep-ex/0304040.

\bibitem{clas}

S. Stepanyan et al., CLAS collaboration, hep-ex/0307018 (to be published in Phys. Rev. Lett.).

\bibitem{saphir}

J. Barth et al., SAPHIR collaboration, Phys. Lett. {\bfseries B572}, 127 (2003); hep-ex/0307083.

\bibitem{nu}

A. E. Asratyan, A. G. Dolgolenko, and M. A. Kubantsev, hep-ex/0309042 (submitted to Yad. Fiz.).

\bibitem{psa}

R. A. Arndt, I. I. Strakovsky, and R. L. Workman, Phys. Rev. {\bfseries C68}, 042201 (2003).

\bibitem{isot}

S. Capstick, P. R. Page, and W. Roberts, Phys. Lett. {\bfseries B570}, 185 (2003);

P. Page, hep-ph/0310200.

\bibitem{iso0}

K. Hicks, private communications.

\bibitem{qn1}

T. Hyodo, A. Hosaka, and E. Oset, nucl-th/0307105.

\bibitem{qn2}

Q. Zhao, hep-ph/0310350.

\bibitem{qn3}

K. Nakayama and K. Tsushima, hep-ph/0311112.

\bibitem{jw}

R. Jaffe and F. Wilczek, Phys. Rev. Lett. {\bfseries 91}, 232003 (2003).

\bibitem{kl}

M. Karliner and H. J. Lipkin, Phys. Lett. {\bfseries B575}, 249 (2003).

\bibitem{exi}

C. Alt, et al., hep-ex/0310014.

\bibitem{dp}

D. Diakonov and V. Petrov, hep-ph/0310212.

\bibitem{ds}

$D_{sJ}^*$(2317), $D_{sJ}$(2458): B. Aubert et al., Phys. Rev. Lett. {\bfseries 90}, 242001 (2003); 

D.Besson et al., Phys. Rev. {\bfseries D68}, 032002 (2003); 

K. Abe et al., hep-ex/0307041;

B. Aubert et al., hep-ex/0310050,

M(3872): K. Abe et al., hep-ex/0308029; 

D. Acosta, hep-ex/0312021.


\end{thebibliography}
\end{document}